\documentclass[sigconf,nonacm]{acmart}
\makeatletter
\@ACM@balancefalse
\makeatother
\AtBeginDocument{%
  }

\usepackage{amsmath}
\usepackage{booktabs}
\usepackage{multirow}
\usepackage{array}
\usepackage{graphicx}
\usepackage{xcolor}
\usepackage{listings}
\usepackage{url}
\usepackage{xurl}
\hypersetup{
  hidelinks,
  pdftitle={CircuitWeave: Topology--Behavior Alignment for Executable Multimodal RTL Generation},
  pdfauthor={Jiahao Feng, Haiyan Qin, Zhiwei Xie, Wang Kang},
  pdfkeywords={RTL generation, electronic design automation, Verilog, multimodal learning, circuit contracts}
}

\definecolor{aspdacblue}{HTML}{2F5597}
\definecolor{aspdacgreen}{HTML}{2E7D6D}
\definecolor{aspdacorange}{HTML}{B45F06}
\definecolor{aspdacred}{HTML}{A94442}
\definecolor{aspdacline}{HTML}{4B5563}
\definecolor{aspdacgray}{HTML}{F5F6F8}
\definecolor{aspdacbluebg}{HTML}{EAF1FB}
\definecolor{aspdacgreenbg}{HTML}{EAF6F2}
\definecolor{aspdacorangebg}{HTML}{FFF4E6}
\definecolor{aspdacredbg}{HTML}{FBEAEA}

\newcommand{\toolname}{\textsc{CircuitWeave}}
\newcommand{\qwen}{Qwen3.5-4B}
\newcommand{\verilog}{Verilog}
\providecommand{\Description}[1]{}

\lstdefinelanguage{Verilog}{
  morekeywords={module,endmodule,input,output,inout,wire,reg,logic,always,always_comb,always_ff,assign,begin,end,if,else,case,endcase,default,posedge,negedge,parameter,localparam,generate,endgenerate,for,integer},
  sensitive=true,
  morecomment=[l]{//},
  morecomment=[s]{/*}{*/},
  morestring=[b]"
}
\title[CircuitWeave]{\texorpdfstring{%
  \vspace*{-3.2em}%
  \parbox{\textwidth}{\centering
    \resizebox{\textwidth}{!}{\normalfont\bfseries\itshape
      This work has been submitted to the IEEE for possible publication. Copyright may be transferred without notice, after which this version may no longer be accessible.}}\\[2.2em]
  \toolname{}: Topology--Behavior Alignment for Executable Multimodal RTL Generation}%
  {CircuitWeave: Topology--Behavior Alignment for Executable Multimodal RTL Generation}}

\author{Jiahao Feng}
\affiliation{%
  \institution{Hangzhou International Innovation Institute, Beihang University}
  \city{Hangzhou}
  \country{China}}
\email{fengjiahao@buaa.edu.cn}

\author{Haiyan Qin}
\affiliation{%
  \institution{National College for Excellent Engineers, Beihang University}
  \city{Beijing}
  \country{China}}
\affiliation{%
  \institution{Jingjinji National Center of Technology Innovation}
  \city{Beijing}
  \country{China}}
\email{haiyanq@buaa.edu.cn}

\author{Zhiwei Xie}
\affiliation{%
  \institution{Hangzhou International Innovation Institute, Beihang University}
  \city{Hangzhou}
  \country{China}}
\email{xiezhiwei@buaa.edu.cn}

\author{Wang Kang}
\authornote{Corresponding author: Wang Kang (wkang@buaa.edu.cn).}
\affiliation{%
  \institution{Hangzhou International Innovation Institute, Beihang University}
  \city{Hangzhou}
  \country{China}}
\affiliation{%
  \institution{School of Integrated Circuit Science and Engineering, Beihang University}
  \city{Beijing}
  \country{China}}
\email{wkang@buaa.edu.cn}

\begin{document}

\begin{abstract}
Text-only LLMs generate RTL from natural-language specifications, but prose can leave connectivity, register boundaries, and state--output relations implicit even when interfaces and cycle-level behavior are specified. Schematics can make these structural relations explicit and thereby complement the behavioral constraints conveyed by text. Yet simply adding an image creates a fusion challenge: direct multimodal decoding does not explicitly separate the evidence roles of text and schematics or make missing and conflicting constraints explicit before code generation. We present \toolname{}, a contract-mediated multimodal framework that extracts a topology contract from the schematic and a behavior contract from the text. It fuses these records into a circuit contract that serializes correspondences, missing evidence, and conflicts, then generates RTL only from this contract. A joint objective supervises both contracts, serialized fusion, contract-conditioned RTL generation, and reverse reconstruction of covered contract fields from reference RTL\@. We construct 5,000 executable-qualified packages, each containing a specification, generated schematic, structured contracts, reference RTL, and self-checking testbench, and use the training split to adapt \qwen{} with LoRA\@. On VerilogEval-Human, \toolname{} reaches 46.60\% pass@1, 61.49\% pass@5, and 65.39\% pass@10. These point estimates are 8.46, 5.85, and 2.57 points above those of the same adapted checkpoint without the schematic. On RTLLM, it reaches 40.00\%, 48.00\%, and 52.00\%, two points above the adapted text-only condition at each cutoff. The dataset is publicly available at \url{https://huggingface.co/datasets/fengjiahao0421/CircuitWeave}.
\end{abstract}



\maketitle
\hypersetup{
  pdftitle={CircuitWeave: Topology--Behavior Alignment for Executable Multimodal RTL Generation},
  pdfauthor={Jiahao Feng, Haiyan Qin, Zhiwei Xie, Wang Kang},
  pdfkeywords={RTL generation, electronic design automation, Verilog, multimodal learning, circuit contracts}
}

\begin{figure*}[!t]
  \centering
  \includegraphics[width=\textwidth]{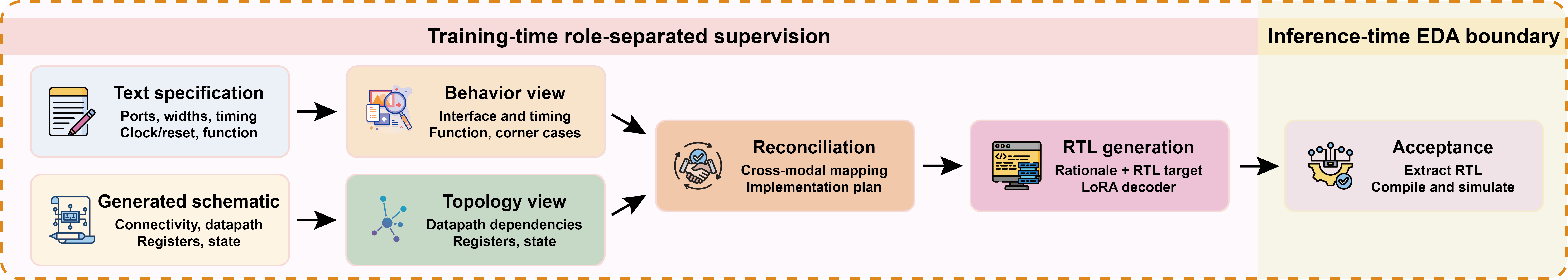}
  \caption{Conceptual overview of \toolname{}. Text and the generated schematic provide behavior and topology evidence, which reconciliation combines into fused contract $z$. The RTL-generation call receives only $z$, and only extracted RTL reaches the hidden EDA boundary. The source-restricted calls follow Eq.~\ref{eq:generation_factorization}; the training-only reverse-reconstruction task is omitted.}
  \Description{High-level CircuitWeave diagram. Text supplies a behavior representation, and the generated schematic supplies a topology representation. Reconciliation combines them into a fused circuit contract before an RTL call conditioned only on that contract. Extracted RTL then reaches hidden executable acceptance. The training-only reverse-reconstruction task is not shown.}
  \label{fig:overview}
\end{figure*}

\section{Introduction}
\label{sec:introduction}
Prior LLM-based RTL methods include English-to-\verilog{} translation, domain adaptation, and hierarchical instruction construction~\cite{pearce2020dave,thakur2024large,zhao2024codev}, as well as structured training tasks and reasoning with executable feedback~\cite{liu2025craftrtl,zhu2025codevr1}. Executable evaluation is stricter than syntax generation: a candidate must satisfy the interfaces, widths, clock/reset semantics, and cycle-level behaviors exercised by a functional testbench. VerilogEval and RTLLM evaluate candidates through compilation and simulation~\cite{liu2023verilogeval,lu2023rtllm}. Intended circuit organization remains a separate structural target because black-box simulation need not verify it.

Text and schematics provide complementary views of hardware intent. Text can specify ports, bit widths, reset polarity, clocks, latency, corner cases, and expected behavior, but may leave connectivity and state organization implicit. Schematics can expose hierarchy, datapaths, register boundaries, and state transitions, but may omit exact interface and timing details. Chang et al.\ report benchmark cases in which diagrams provide hardware information absent from language~\cite{chang2024natural}. Thus, for independently authored specifications, neither modality fully substitutes for the other. In our controlled benchmark study, however, each schematic is generated from the same description and provides a structured visual restatement rather than an independent specification.

These roles motivate a more precise question than whether an image helps: under what training conditions does a description-derived schematic improve executable RTL generation? A single text+image score cannot distinguish RTL-specific adaptation from image conditioning. Without an explicit division of evidence, direct fusion need not preserve which constraints come from text and which come from the schematic. A model may follow visual connectivity while violating a textual width, or match the interface while inserting an incorrect register boundary. We therefore encode the two evidence roles separately and vary checkpoint state and image input in a matched design.

We formulate this setting as topology--behavior contract mediation and instantiate it in \toolname{}. The topology contract $z_v$ records components, connectivity, datapath operators, and sequential structure from the schematic. The behavior contract $z_t$ records interfaces, widths, clock/reset semantics, latency, corner cases, and required function from the text. The fusion stage serializes correspondences, missing evidence, and conflicts in a circuit contract $z$. A separate stage generates RTL conditioned on $z$; the raw modalities are excluded from the RTL-generation call. Figure~\ref{fig:overview} gives the conceptual overview. Equation~\ref{eq:generation_factorization} and Listing~\ref{lst:prompt_template} specify the source-restricted calls and fused-contract-only RTL stage. We jointly optimize contract extraction, serialized fusion, contract-conditioned RTL generation, and reverse reconstruction of covered contract fields from reference RTL, and adapt \qwen{} using LoRA~\cite{qwen2026qwen35,hu2021lora}.

We evaluate this setting using 5,000 executable-qualified packages and a matched $2\times2$ comparison of checkpoint state and input condition. On VerilogEval-Human, the adapted text+image condition reaches 46.60\% pass@1, 61.49\% pass@5, and 65.39\% pass@10. At pass@1, its image-input difference is $+8.46$ points, compared with $-3.01$ points for the base checkpoint. On RTLLM, the adapted text+image condition reaches 40.00\%, 48.00\%, and 52.00\%, two points above the adapted text-only condition at each cutoff. Across these descriptive point estimates, the image-input difference varies by checkpoint state, benchmark, and sampling cutoff (Tables~\ref{tab:benchmark_comparison} and~\ref{tab:image_interaction}).

This work makes three main contributions:
\begingroup
\begin{itemize}
\setlength{\itemsep}{3pt}
  \item We introduce contract-mediated multimodal RTL generation, which constructs modality-specific topology and behavior contracts, fuses them into a circuit contract, and decodes RTL from that fused record. The joint objective supervises contract extraction, serialized fusion, RTL generation, and reverse reconstruction of covered contract fields.
  \item We construct 5,000 executable-qualified design--verification packages through a role-specialized multi-agent pipeline for schematic generation, contract construction, and testbench synthesis. Qualification requires the reference RTL/\allowbreak testbench pair to elaborate, terminate, and satisfy the expected checks under Icarus.
  \item We conduct a matched $2\times2$ comparison of checkpoint adaptation and generated-schematic input. On VerilogEval-Human pass@1, the image-input difference is $-3.01$ points for the base checkpoint and $+8.46$ points for the adapted checkpoint; interactions at larger cutoffs and on RTLLM are smaller or mixed.
\end{itemize}
\endgroup

\section{Related Work}
\label{sec:related_work}
We compare \toolname{} with prior work on language-based RTL generation, multimodal generation, structured circuit representations, and executable verification.

\subsection{Language-Based RTL Generation}
Language-based RTL systems span English-to-\verilog{} translation, domain adaptation, and reasoning-enhanced generation. DAVE studied English-to-RTL translation~\cite{pearce2020dave}; VeriGen studied domain adaptation~\cite{thakur2024large}; and CodeV constructed hierarchical instruction data from RTL repositories~\cite{zhao2024codev}. VeriThoughts combines reasoning data with formal-verification labels~\cite{yubeaton2025verithoughts}; ReasoningV uses adaptive hybrid reasoning~\cite{qin2025reasoningv}; and VeriReason, VeriRL, and CodeV-R1 use simulator feedback or reinforcement learning~\cite{wang2025verireason,teng2025verirl,zhu2025codevr1}. These systems operate on language, code, reasoning, or EDA traces. \toolname{} instead inserts a specification-side circuit contract before RTL generation.

\subsection{Multimodal RTL Generation}
Hardware specifications use circuit, timing, and FSM diagrams to encode relations that can be cumbersome to state in prose. Chang et al.\ report cases in which language omits information needed to recover the intended hardware~\cite{chang2024natural}. VeriGround reports that blank-image and anonymized-identifier controls can reveal circuit-to-RTL improvements without corresponding visual grounding~\cite{yang2026veriground}. \toolname{} instead maps each modality to a source-specific contract and uses matched checkpoint-by-input conditions. Because the present study lacks these grounding controls, its image deltas measure conditioning rather than verified visual grounding.

\subsection{Structured Circuit Representations}
Prior structure-aware methods constrain hardware generation without image input. CraftRTL uses correct-by-construction non-image tasks~\cite{liu2025craftrtl}, whereas CPPL lowers a typed circuit IR under legality checks~\cite{yin2026cppl}. \toolname{} does not replace RTL with a compiler IR\@. Its specification-side contracts retain separate schematic-topology and textual-behavior records before fusion; the fused record then conditions RTL generation.

\subsection{Executable Verification and Adaptation}
Executable tests are a standard correctness signal in code generation~\cite{chen2021evaluating}, while RTL testbenches can check top-level interfaces and the cycle-level behavior they cover. VerilogEval and RTLLM evaluate generated RTL by simulation~\cite{liu2023verilogeval,lu2023rtllm}, and LLM4DV generates verification stimuli~\cite{zhang2023llm4dv}. VeriReason and VeriRL use testbench or simulation feedback~\cite{wang2025verireason,teng2025verirl}; QiMeng-SALV introduces signal-aware learning~\cite{huang2025qimengsalv}; and CodeV-R1 combines reasoning supervision with reinforcement learning~\cite{zhu2025codevr1}. Before training, \toolname{} executes each reference RTL/testbench pair; qualification means that the pair elaborates, terminates, and satisfies the expected checks. During benchmark evaluation, generated RTL is accepted only if it compiles and passes simulation against the hidden testbench.

\section{CircuitWeave: Contract-Mediated Multimodal RTL Generation}
\label{sec:method}
\toolname{} decomposes the pipeline into contract construction, contract fusion, and RTL generation with executable acceptance. Separate source-restricted calls predict a topology contract from the schematic and a behavior contract from the text. Their fused contract is the only serialized input supplied to the RTL-generation call. Training combines losses for the two source-specific contracts, serialized fusion, RTL generation, and reverse RTL-to-contract reconstruction. Figure~\ref{fig:overview} summarizes this evidence flow. The following subsections define the source-restricted calls, fused-only RTL decoder, training objective, and executable boundary.

\subsection{Problem Formulation}
Let $x_t$ denote a textual specification and $x_v$ a generated schematic. Let $z_t$ and $z_v$ denote the behavior and topology contracts, respectively, and let $z$ be the fused circuit contract decoded from $p_{\theta}(z\mid z_v,z_t)$. Finally, let $y$ denote the serialized RTL response containing the generated \verilog{} module. We parameterize the conditional model according to the following source-restricted inference graph:
\begin{equation}
\label{eq:generation_factorization}
\begin{aligned}
&p_{\theta}(z_v,z_t,z,y\mid x_v,x_t) \\
&=p_{\theta}(z_v\mid x_v)\,
  p_{\theta}(z_t\mid x_t)\,
  p_{\theta}(z\mid z_v,z_t)\,
  p_{\theta}(y\mid z).
\end{aligned}
\end{equation}
Equation~\ref{eq:generation_factorization} encodes the designed conditional dependencies among evidence extraction, contract fusion, and code generation. Under this parameterization, marginalizing the intermediate contract sequences gives
\begin{equation}
\label{eq:contract_marginal}
\begin{aligned}
p_{\theta}(y\mid x_t,x_v)=\sum_{z_v,z_t,z}
&p_{\theta}(y\mid z)p_{\theta}(z\mid z_v,z_t)\\[-2pt]
&\cdot p_{\theta}(z_v\mid x_v)p_{\theta}(z_t\mid x_t).
\end{aligned}
\end{equation}
At inference, we do not enumerate all intermediate contracts in Equation~\ref{eq:contract_marginal}. Instead, we follow the factorization through two phases and four separate decoder calls. In the contract phase, the topology call receives only $x_v$, the behavior call receives only $x_t$, and the fusion call receives only $z_v,z_t$. In the RTL phase, the fourth call receives only serialized $z$ and emits $y$. Thus, the raw text and image are not supplied directly to the RTL-generation call. Source restriction denotes this call-level input constraint, not statistical independence between the modalities. When the image is absent, $z_v$ contains explicit \texttt{unknown} fields instead of inferred visual evidence.

\subsection{Contract-Annotated Executable Package Construction}
Each of 5,000 design--verification packages contains the five artifact classes in Table~\ref{tab:package_artifacts}. The contract record contains separate $z_v$, $z_t$, and $z$ sections; the testbench remains a hidden qualification artifact and is never a model input.

\begin{table}[t]
  \centering
  \caption{Artifact source and role in each package. Contracts supervise generation; only the reference RTL/testbench pair is EDA-executed.}
  \label{tab:package_artifacts}
  \footnotesize
  \setlength{\tabcolsep}{4.0pt}
  \renewcommand{\arraystretch}{1.10}
  \begin{tabular}{@{}>{\raggedright\arraybackslash}p{0.19\columnwidth}
                          >{\raggedright\arraybackslash}p{0.24\columnwidth}
                          >{\raggedright\arraybackslash}p{0.47\columnwidth}@{}}
    \toprule
    Artifact & Construction source & Learning/validation role \\
    \midrule
    Text & Web collection & Source for behavior contract $z_t$ \\
    Schematic & Schematic agent & Source for topology contract $z_v$ \\
    Contract record & Contract agent & Targets $z_v$, $z_t$, and fused $z$ \\
    RTL & Web collection & Target $y$; compiled and simulated \\
    Testbench & Verification agent & Hidden qualifier; executed with RTL \\
    \bottomrule
  \end{tabular}
\end{table}

We construct each design--verification package through a role-specialized multi-agent pipeline. Starting from a publicly accessible \verilog{} module and its natural-language description, a schematic agent translates the design intent into a visual representation. A contract agent then receives only its designated input in each pass: the topology pass receives the schematic, the behavior pass receives the description, and the fusion pass receives the two resulting contracts. During contract construction, this agent receives neither the reference RTL nor the testbench. These call-level restrictions remove the agent's direct access to the reference artifacts. In parallel, a verification agent synthesizes a self-checking testbench from the textual specification. An executable validator finally compiles and simulates the reference RTL/testbench pair, retaining only packages that elaborate, terminate, and satisfy the expected checks.

Nano Banana 2, Gemini 3 Flash, and GPT-5.5 instantiate the schematic, contract, and verification roles, respectively~\cite{google2026nanobanana2,google2026gemini3flash,openai2026gpt55}. The executable validator uses Icarus Verilog, invoking \texttt{iverilog} for compilation and elaboration and \texttt{vvp} for simulation~\cite{icarusverilog}.

This executable gate retains reference RTL/testbench pairs that elaborate and pass simulation; it does not establish test coverage, cross-modal fidelity, or semantic correctness beyond the exercised checks. The fusion and reverse-reconstruction losses are training signals rather than certification. We therefore call the retained packages \emph{executable-qualified}. After normalizing module identifiers, contract headings, code fences, answer tags, and simulator outcomes, we screen within-corpus duplicates and split 4,750/250 packages for training/validation. The available records do not support a benchmark-aware semantic-overlap count, which we treat as a validity threat in Section~\ref{sec:limitations}.

Table~\ref{tab:training_dataset_comparison} compares this package design with representative RTL-training resources. Each package in this work co-locates a generated schematic, modality-specific and fused contracts, reference RTL, and an executable testbench.

\begin{table*}[t]
  \centering
  \caption{Representative RTL-generation training resources. ``TB use'' describes testbenches used in curation or learning, not necessarily model inputs. Scale is rounded from the cited releases.}
  \label{tab:training_dataset_comparison}
  \footnotesize
  \setlength{\tabcolsep}{3.2pt}
  \renewcommand{\arraystretch}{1.08}
  \begin{tabular}{@{}>{\raggedright\arraybackslash}p{0.16\textwidth}
                          >{\centering\arraybackslash}p{0.08\textwidth}
                          >{\raggedright\arraybackslash}p{0.25\textwidth}
                          >{\centering\arraybackslash}p{0.06\textwidth}
                          >{\centering\arraybackslash}p{0.08\textwidth}
                          >{\raggedright\arraybackslash}p{0.26\textwidth}@{}}
    \toprule
    Resource & Scale & Learning target & Image & TB use & Pair-level quality control \\
    \midrule
    RTLCoder-27K~\cite{liu2023rtlcoder} & 27K & Instruction + RTL SFT & No & No & Syntax checking \\
    CodeV-Verilog~\cite{zhao2024codev} & 165K & Multilevel summaries + RTL SFT & No & No & Syntax, deduplication, and decontamination \\
    CraftRTL~\cite{liu2025craftrtl} & 110K & Structured tasks + RTL/repair & No$^{\dagger}$ & No & Construction checks and filtered repair \\
    OriGen~\cite{cui2024origen} & 222K & RTL augmentation + repair & No & No & Compiler-guided syntax repair \\
    VeriThoughts~\cite{yubeaton2025verithoughts} & 20K & Reasoning + RTL SFT & No & No & Formal-equivalence labels \\
    CodeV-R1~\cite{zhu2025codevr1} & 87K+3.1K & Reasoning SFT + RL & No & Filter/RL & Generated-testbench equivalence \\
    Veribench-53K~\cite{teng2025verirl} & 53K & Reward modeling + RL & No & 5--10/pair & Multi-testbench simulation \\
    \midrule
    \textbf{This work} & \textbf{5K} & \textbf{Topology/behavior/fused contracts + RTL} & \textbf{Yes} & \textbf{1/pair} & \textbf{Icarus compilation and simulation} \\
    \bottomrule
  \end{tabular}
  \vspace{2pt}
  \parbox{0.98\textwidth}{\footnotesize $^{\dagger}$CraftRTL includes K-maps, FSMs, and waveforms as structured non-image representations. Executable QC does not imply full semantic certification.}
\end{table*}

\subsection{Topology--Behavior Contract Fusion}
The topology contract $z_v$ assigns the schematic a structural role. Its fixed headings cover components and ports, connectivity, datapath operators, register boundaries, state transitions, internal dependencies, and \texttt{unknown} evidence. The behavior contract $z_t$ assigns the text a complementary role and records the module interface, widths, clock/reset semantics, function, latency, output timing, and corner cases. Here, source isolation is a call-level restriction: the topology pass receives only $x_v$, and the behavior pass receives only $x_t$. It removes direct access to the other raw input before fusion but does not imply provenance or statistical independence.

The fusion target aligns shared entities, carries forward modality-specific constraints, and labels shared fields as supported, missing, or conflicting. Explicit textual interface and timing requirements fill fields absent from the schematic, while visible structural relations retain their $z_v$ provenance. The resulting $z$ serializes the cross-modal correspondence and the constraints supplied to the RTL-generation call. Listing~\ref{lst:prompt_template} shows the two-phase serialization used by the model.

Table~\ref{tab:contract_fields} summarizes the contract groups and their fusion policy. The serialized schema stores each field value with its evidence source. In the target records, unsupported fields are labeled \texttt{unknown}. When two explicit values disagree, the field remains \texttt{conflict} unless a stated field-specific authority rule applies; any unresolved conflict remains explicit in $z$. These labels distinguish missing from conflicting evidence instead of leaving both implicit in a free-form rationale.

\begin{table}[t]
  \centering
  \caption{Contract fields and their treatment in fused $z$ and reverse-reconstruction target $z_{\mathrm{cov}}$.}
  \label{tab:contract_fields}
  \footnotesize
  \setlength{\tabcolsep}{2.6pt}
  \renewcommand{\arraystretch}{1.08}
  \begin{tabular}{@{}>{\raggedright\arraybackslash}p{0.16\columnwidth}
                          >{\raggedright\arraybackslash}p{0.22\columnwidth}
                          >{\raggedright\arraybackslash}p{0.24\columnwidth}
                          >{\raggedright\arraybackslash}p{0.28\columnwidth}@{}}
    \toprule
    Group & Topology $z_v$ & Behavior $z_t$ & Treatment in $z$ / $z_{\mathrm{cov}}$ \\
    \midrule
    Interface & Visible pins and links & Name, direction, width & $z$: fuse ports; $z_{\mathrm{cov}}$: declarations \\
    Sequential & Registers, state edges & Clock, reset, latency & $z$: fuse state/timing; $z_{\mathrm{cov}}$: event controls \\
    Datapath & Operators, dependencies & Function, corner cases & $z$: fuse operations; $z_{\mathrm{cov}}$: selected dependencies \\
    Status & Visible or \texttt{unknown} & Stated or \texttt{unknown} & $z$: preserve source/status; $z_{\mathrm{cov}}$: omit \texttt{unknown}/\texttt{conflict} \\
    \bottomrule
  \end{tabular}
\end{table}

\begin{lstlisting}[float=t,language={},caption={Source-restricted contract calls and fused-only RTL call.},label={lst:prompt_template}]
Contract phase:
A. schematic only -> <topology> components; edges; operators; state </topology>
B. text only -> <behavior> ports/widths; clock/reset; timing; function </behavior>
C. z_v and z_t only -> <fused> correspondence; missing/conflict; RTL constraints </fused>
RTL phase:
D. fused z only -> <answer> complete Verilog module </answer>
\end{lstlisting}

\subsection{Contract-Conditioned RTL Generation}
We train the shared decoder on five conditional tasks with length-normalized token losses. Let $z_{\mathrm{cov}}=P_{\mathrm{cov}}(z)$ be a deterministic projection of $z$. It retains port declarations, register/state elements, clock/reset event controls, and explicit datapath operators and dependencies. The projection omits \texttt{unknown}/\texttt{conflict} values and behavioral requirements outside the scope of the reverse task, including latency, corner-case behavior, and full functional equivalence. The losses $\mathcal{L}_{\mathrm{topo}}$, $\mathcal{L}_{\mathrm{behavior}}$, and $\mathcal{L}_{\mathrm{RTL}}$ predict $z_v$ from $x_v$, $z_t$ from $x_t$, and reference RTL $y$ from fused $z$, respectively. The serialized fusion loss $\mathcal{L}_{\mathrm{align}}=-\log p_{\theta}(z\mid z_v,z_t)$ supervises the target fused record, including its correspondence and status fields. The reverse-reconstruction loss $\mathcal{L}_{\mathrm{cons}}=-\log p_{\theta}(z_{\mathrm{cov}}\mid y)$ predicts $z_{\mathrm{cov}}$ from reference RTL and supplies an indirect auxiliary signal through the shared parameters. The complete objective is
\begin{equation}
\label{eq:joint_objective}
\mathcal{L}=
\mathcal{L}_{\mathrm{topo}}+
\mathcal{L}_{\mathrm{behavior}}+
\mathcal{L}_{\mathrm{RTL}}+
\lambda\mathcal{L}_{\mathrm{align}}+
\mu\mathcal{L}_{\mathrm{cons}}.
\end{equation}
Because $z$ explicitly serializes field correspondences and their status, $\mathcal{L}_{\mathrm{align}}$ supervises those targets but is not an independent measurement of alignment accuracy. We use LoRA for parameter-efficient adaptation of the base multimodal model. The training scheme combines contract-mediated generation with a reverse-reconstruction auxiliary objective. Concretely, trainable low-rank updates are inserted into the \qwen{} language-model linear layers while the base weights, vision encoder, and multimodal aligner remain frozen~\cite{hu2021lora,msswift}. No benchmark testbench or simulator outcome is exposed during training.

Each term has a teacher-forced gradient path to the shared decoder parameters. The contract-phase terms supervise the two source-specific serializations and their fusion; the RTL term supervises $z\rightarrow y$; and the reverse term supervises $y\rightarrow z_{\mathrm{cov}}$. Training supplies the reference fused contract to the RTL task, whereas inference supplies the predicted contract from the contract phase. Because the reverse task does not compare a sampled $\hat y$ with $z$, it regularizes the shared parameters only indirectly and does not enforce per-candidate contract--RTL consistency. It is neither an EDA reward nor a claim of formal equivalence.

\subsection{Candidate Extraction and Executable Acceptance}
All base and adapted variants receive the same two-stage and output-format instructions. Let $E(y)$ extract a complete module from the final \texttt{<answer>} block and return $\bot$ otherwise. The evaluator removes response markup but does not repair code. It compiles $E(y)$ with hidden testbench $T$ and simulates each successfully elaborated candidate:
\begin{equation}
\label{eq:acceptance_indicator}
\begin{aligned}
A(y,T)=\mathbf{1}\!\bigl[&E(y)\ne\bot\ \land
\operatorname{compile}(E(y),T)\\[-2pt]
&{}\land\operatorname{simulate}(E(y),T)\bigr].
\end{aligned}
\end{equation}
Extraction failures, interface mismatches, compilation errors, timeouts, and testbench-detected functional mismatches all receive zero. The reverse auxiliary task reconstructs selected contract fields from reference RTL during training, whereas the hidden testbench defines the final benchmark acceptance decision for generated candidates. A passing result establishes only satisfaction of the behavior exercised by $T$. Every condition uses the same extraction and EDA procedure without checker-based reranking or post-generation repair.

\section{Experimental Setup}
\label{sec:experiments}
The experiments cross checkpoint state with inference condition in a matched $2\times2$ design. \textbf{RQ1} quantifies the end-to-end difference between the base checkpoint and the checkpoint adapted with the complete contract-mediated objective. \textbf{RQ2} quantifies the within-checkpoint difference associated with adding a description-derived schematic. \textbf{RQ3} examines whether the resulting descriptive interaction has a consistent direction across benchmarks and sampling cutoffs.

\begin{table*}[t]
\centering
\caption{Executable RTL-generation results (\%). Controlled rows share decoding and EDA settings; bold marks the best controlled value for each metric. Reported external rows provide scale context only because protocols differ.}
\label{tab:benchmark_comparison}
\small
\renewcommand{\arraystretch}{1.08}
\setlength{\tabcolsep}{5pt}
\begin{tabular}{@{}>{\raggedright\arraybackslash}p{0.10\textwidth}
                    >{\raggedright\arraybackslash}p{0.26\textwidth}
                    >{\raggedright\arraybackslash}p{0.10\textwidth}
                    cccccc@{}}
\toprule
\multirow{2}{*}{Group} & \multirow{2}{*}{Model} & \multirow{2}{*}{Input} &
\multicolumn{3}{c}{VerilogEval-Human} & \multicolumn{3}{c}{RTLLM} \\
\cmidrule(lr){4-6}\cmidrule(lr){7-9}
 & & & pass@1 & pass@5 & pass@10 & pass@1 & pass@5 & pass@10 \\
\midrule
Controlled & Base \qwen{} & Text & 23.59 & 41.41 & 48.08 & 28.00 & 35.00 & 41.00 \\
Controlled & Base \qwen{} & Text+image & 20.58 & 42.97 & 51.28 & 27.00 & 39.00 & 44.00 \\
Controlled & \toolname{} w/o image & Text & 38.14 & 55.64 & 62.82 & 38.00 & 46.00 & 50.00 \\
Controlled & \textbf{CircuitWeave} & Text+image & \textbf{46.60} & \textbf{61.49} & \textbf{65.39} & \textbf{40.00} & \textbf{48.00} & \textbf{52.00} \\
\midrule
Reported & GPT-4o~\cite{liu2025craftrtl} & Text & 57.1 & 63.9 & 66.7 & 33.8 & 44.4 & 48.3 \\
Reported & DeepSeek-V3~\cite{huang2025qimengsalv} & Text & 70.7 & 77.4 & 78.8 & 62.0 & 72.0 & 72.4 \\
Reported & StarCoder2-15B~\cite{liu2025craftrtl} & Text & 37.7 & 50.6 & 57.2 & 15.5 & 37.6 & 45.7 \\
Reported & CodeV-CodeQwen~\cite{liu2025craftrtl} & Text & 53.2 & 65.1 & 68.5 & 36.6 & 53.3 & 61.3 \\
Reported & OriGen~\cite{huang2025qimengsalv} & Text & 54.4 & 60.1 & 64.2 & 50.6 & 68.3 & 74.3 \\
Reported & CodeV-R1-7B~\cite{zhu2025codevr1} & Text & 69.9 & 79.3 & 81.7 & 72.9 & 86.1 & -- \\
\bottomrule
\end{tabular}
\end{table*}

\subsection{Benchmarks, Conditions, and Metric}
We evaluate VerilogEval-Human (156 tasks)~\cite{liu2023verilogeval} and RTLLM~\cite{lu2023rtllm}, both of which provide textual tasks and functional testbenches. For each task and condition, we draw 20 samples at temperature 0.7 and compile and simulate every candidate independently. We report pass@1, pass@5, and pass@10 using the standard unbiased estimator of Chen et al.~\cite{chen2021evaluating}, averaged across tasks without category reweighting. A candidate must satisfy Eq.~\ref{eq:acceptance_indicator}; similarity to reference RTL is not used.

The four conditions in Table~\ref{tab:benchmark_comparison} cross checkpoint state (base or adapted) with contract input (text or text+image). In the text+image condition, Nano Banana 2 generates a schematic from the released task description alone and does not receive the reference RTL or benchmark testbench. In text-only inference, the topology contract uses explicit \texttt{unknown} values; in text+image inference, it is extracted from the generated schematic. The same LoRA adapter is evaluated in both modes. Every row follows the same contract-first, RTL-second protocol, sampling budget, extraction procedure, and EDA harness, without reranking or repair.

\subsection{Training and Comparison Protocol}
We train on 4,750 packages and validate on 250. Training optimizes Eq.~\ref{eq:joint_objective}, with each term normalized by its supervised token count. Separate source-restricted examples implement the two contract branches, and the reference fused contract conditions $\mathcal{L}_{\mathrm{RTL}}$ during teacher forcing. The base checkpoint is \qwen{}~\cite{qwen2026qwen35}. Training uses bfloat16 on one GPU and the ms-swift \texttt{all-linear} LoRA target~\cite{msswift}. Rank is 32 with scaling factor 64. We train for three epochs at learning rate $1\times10^{-4}$, warmup ratio 0.05, and sequence length 8192. Per-device batch size is 2 with gradient accumulation 2, and the image-token budget is 1024. Evaluation and checkpointing occur every 100 steps, with two checkpoints retained.

External systems in Table~\ref{tab:benchmark_comparison} differ in prompting, model scale, decoding, benchmark revision, and post-processing. Their reported values provide approximate context for the compact 4B model rather than head-to-head comparisons. All claims about adaptation and image input rely on the four controlled rows.

\section{Results and Analysis}
\label{sec:results}
\subsection{RQ1: Base-to-Adapted Performance}
Across both benchmarks and all three $k$ values, the adapted checkpoint has higher point estimates than the corresponding base checkpoint. At $k=1,5,10$, text-only VerilogEval-Human pass@$k$ rises by 14.55, 14.23, and 14.74 points to 38.14\%, 55.64\%, and 62.82\%. With text+image, the corresponding differences are 26.02, 18.52, and 14.11 points, reaching 46.60\%, 61.49\%, and 65.39\%. On RTLLM, the base-to-adapted differences are 10, 11, and 9 points for text and 13, 9, and 8 points for text+image. These comparisons evaluate the combined adaptation recipe, including all five objective terms and LoRA; they do not identify the contribution of any individual branch, loss, or adaptation component.

\subsection{RQ2--RQ3: Checkpoint--Input Interaction}
The within-checkpoint deltas in Table~\ref{tab:image_interaction} summarize the difference associated with adding the generated schematic. On VerilogEval-Human pass@1, the image delta changes from $-3.01$ points in the base model to $+8.46$ after adaptation, giving a descriptive interaction of 11.47 points. The corresponding interaction is $+4.29$ at pass@5 but $-0.63$ at pass@10. On RTLLM, the adapted image delta is consistently $+2$ points, while the interactions are $+3$, $-2$, and $-1$ points at pass@1, pass@5, and pass@10. The interaction is largest at the lower VerilogEval-Human cutoffs and does not have a consistent direction across benchmarks or sampling cutoffs.

\begin{table}[t]
\centering
\caption{Image-input deltas (generated schematic minus text only, percentage points). Interaction is the adapted delta minus the base delta; all values are descriptive. ``VerilogEval'' denotes VerilogEval-Human.}
\label{tab:image_interaction}
\small
\renewcommand{\arraystretch}{1.12}
\begin{tabular*}{\columnwidth}{@{\extracolsep{\fill}}lcrrr@{}}
\toprule
Benchmark & $k$ & Base $\Delta$ & Adapted $\Delta$ & Interaction \\
\midrule
VerilogEval & 1  & $-3.01$ & $+8.46$ & $+11.47$ \\
            & 5  & $+1.56$ & $+5.85$ & $+4.29$ \\
            & 10 & $+3.20$ & $+2.57$ & $-0.63$ \\
RTLLM      & 1  & $-1.00$ & $+2.00$ & $+3.00$ \\
            & 5  & $+4.00$ & $+2.00$ & $-2.00$ \\
            & 10 & $+3.00$ & $+2.00$ & $-1.00$ \\
\bottomrule
\end{tabular*}
\end{table}

The reported external rows contextualize scale but do not form a unified leaderboard. In the reported values, \toolname{} is above StarCoder2-15B at every VerilogEval-Human cutoff and on all three RTLLM metrics; because the protocols differ, these are contextual rather than head-to-head comparisons. Several external systems also report higher values on parts of the table. Without blank-image, mismatched-image, or identifier-anonymization controls, the positive image-minus-text deltas for the adapted checkpoint are consistent with generated-schematic conditioning but do not verify visual grounding~\cite{yang2026veriground}. Task-level bootstrap intervals and stage-wise failure counts are unavailable, so all differences remain descriptive, with uncertainty especially consequential for the smaller deltas. Under the shared EDA protocol, the adapted checkpoint has positive image-minus-text differences on all reported metrics, whereas evidence for a stable adaptation-by-image interaction is mixed.

\begin{figure}[!t]
  \centering
  \includegraphics[width=\columnwidth]{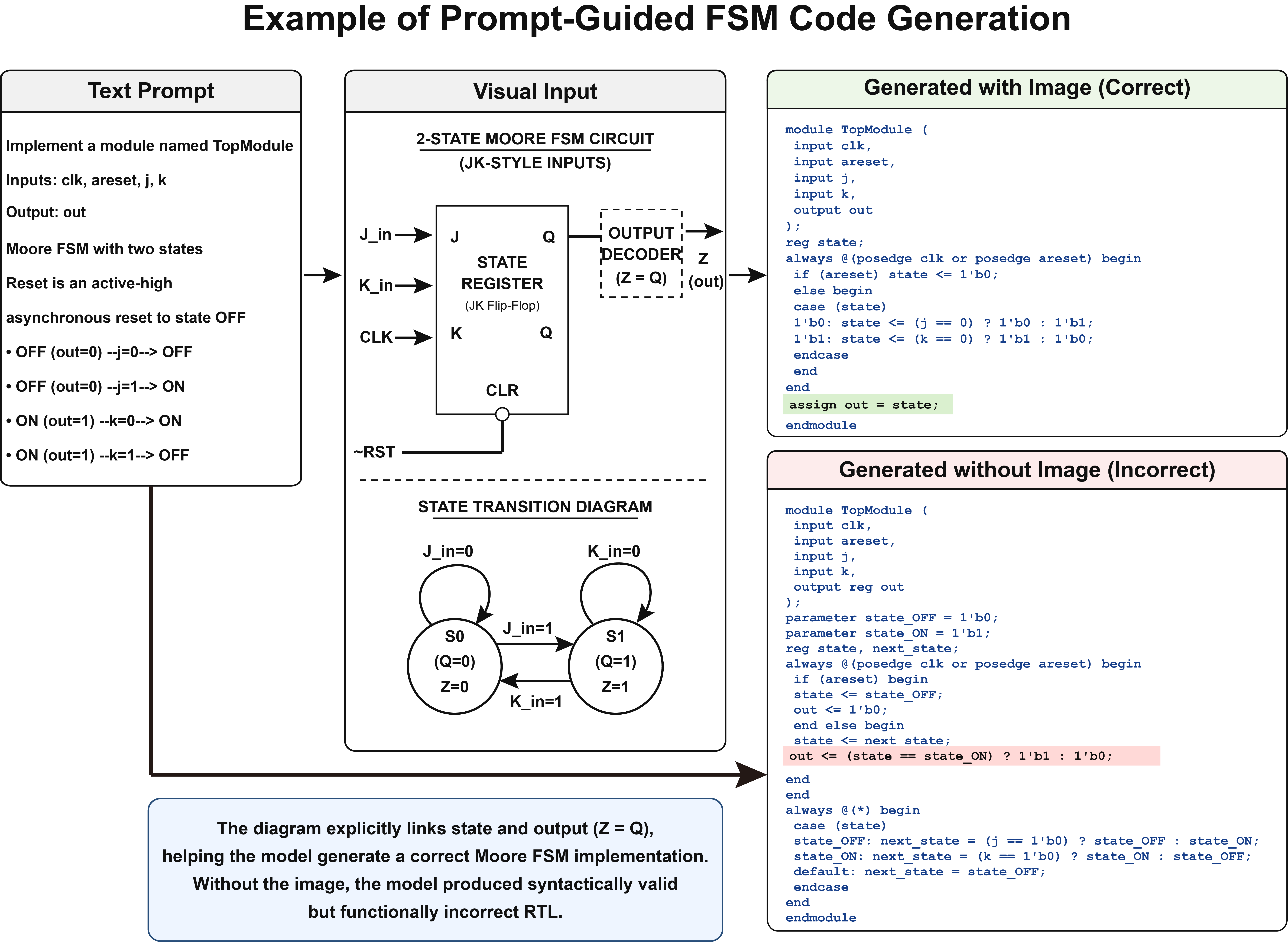}
  \caption{Paired outputs from the same adapted \qwen{} checkpoint on one FSM task. The image-conditioned candidate uses the current state; the text-only candidate uses the pre-transition state and introduces a one-cycle lag. Highlights mark the assignments.}
  \Description{Paired qualitative comparison from the same adapted Qwen3.5-4B checkpoint for the same two-state Moore-FSM text prompt. In the text-plus-image condition, the Verilog candidate continuously assigns out from state. In the text-only condition, the candidate updates out in the clocked block from the old state, causing a one-cycle output delay after each state transition.}
  \label{fig:fsm_case}
\end{figure}

\subsection{Qualitative FSM Case Study}
Figure~\ref{fig:fsm_case} compares outputs from the same adapted \qwen{} checkpoint on the same FSM prompt. The model and text are fixed; one condition also receives the generated schematic. Both candidates recover the reset and transitions. The schematic specifies $Z=Q$, which the image-conditioned candidate implements as \texttt{assign out = state}. The text-only candidate updates \texttt{out} in the clocked block; nonblocking semantics therefore use the pre-edge state and cause a one-cycle lag on both transitions. This paired example illustrates how schematic conditioning can help the same model avoid a concrete RTL error and complements the adapted checkpoint's positive aggregate image-minus-text deltas in Tables~\ref{tab:benchmark_comparison} and~\ref{tab:image_interaction}.

\section{Limitations}
\label{sec:limitations}
The present study is limited to description-derived schematics, one 4B checkpoint, and simulation-level evaluation. Passing Icarus qualification establishes agreement between the reference RTL and testbench only for the exercised behaviors; it does not establish coverage or cross-modal fidelity, and weak tests may accept incorrect RTL\@. Web collection also creates provenance, licensing, and benchmark-overlap risks. The reported results are aggregate point estimates without repeated runs across training or sampling seeds, paired task-level intervals, or stage-wise failure counts. Evaluation excludes synthesis, timing, area, power, and physical design.

The controlled design compares checkpoint state and the addition of a description-derived schematic; it does not isolate the individual contributions of the contract branches, $\mathcal{L}_{\mathrm{align}}$, $\mathcal{L}_{\mathrm{cons}}$, or LoRA\@. Generated contracts have not been independently audited, and the schematics derive from the same text used by the text-only condition. Without blank, mismatched, anonymized, or topology-perturbed image controls, the results quantify conditioning on these generated schematics rather than verified visual grounding. Future evaluation should include objective ablations, independently authored diagrams, broader contract coverage, and coverage-guided tests.

\section{Conclusion}
\label{sec:conclusion}
We presented \toolname{}, which fuses schematic topology and textual behavior into a contract before RTL decoding. LoRA adaptation of \qwen{} uses the training split of a 5,000-package executable-qualified corpus. Under the text+image condition, the adapted model reaches 46.60\% pass@1 on VerilogEval-Human and 40.00\% on RTLLM\@. On VerilogEval-Human pass@1, the image-input difference is $-3.01$ points for the base checkpoint and $+8.46$ points for the adapted checkpoint. Under the shared evaluation protocol, these point estimates characterize the end-to-end performance of contract-mediated multimodal RTL generation; they do not isolate individual adaptation components or establish visual grounding.

\clearpage
\bibliographystyle{ACM-Reference-Format}
\bibliography{references_no_grpo}

\end{document}